\documentclass[runningheads]{llncs}
\usepackage[utf8]{inputenc}
\usepackage{longtable}
\usepackage{supertabular,booktabs}
\usepackage{caption}
\usepackage{multirow}
\usepackage{graphicx}
\usepackage{float}
\usepackage{comment}

\begin{document}

\title{Privacy Concerns in Chatbot Interactions: When to Trust and When to Worry \thanks{This is the author version of the article: Belen Saglam R., Nurse J.R.C., Hodges D. (2021) Privacy Concerns in Chatbot Interactions: When to Trust and When to Worry. In: Stephanidis C., Antona M., Ntoa S. (eds) HCI International 2021. HCII 2021. Communications in Computer and Information Science, vol 1420. Springer, Cham. https://doi.org/10.1007/978-3-030-78642-7{\_}53}}

\titlerunning{Privacy Concerns in Information Disclosure via Chatbots}

\author{Rahime Belen Saglam\inst{1} \and
Jason R.C. Nurse\inst{1} \and
Duncan Hodges\inst{2}}
\authorrunning{Saglam, Nurse and Hodges}

\institute{University of Kent, UK\\ 
\email{\{R.Belen-Saglam-724, J.R.C.Nurse\}@kent.ac.uk}
\and
Cranfield University, Defence Academy of the United Kingdom, UK\\
\email{d.hodges@cranfield.ac.uk}}
\maketitle              
\begin{abstract}
Through advances in their conversational abilities, chatbots have started to request and process an increasing variety of sensitive personal information. The accurate disclosure of sensitive information is essential where it is used to provide advice and support to users in the healthcare and finance sectors.
In this study, we explore users' concerns regarding factors associated with the use of sensitive data by chatbot  providers. We surveyed a representative sample of 491 British citizens. 
Our results show that the user concerns focus on deleting personal information and concerns about their data's inappropriate use. We also identified that individuals were concerned about losing control over their data after a conversation with conversational agents. We found no effect from a user's gender or education but did find an effect from the user's age, with those over 45 being more concerned than those under 45. 
We also considered the factors that engender trust in a chatbot. Our respondents' primary focus was on the chatbot's technical elements, with factors such as the response quality being identified as the most critical factor. We again found no effect from the user's gender or education level; however, when we considered some social factors (e.g. avatars or perceived `friendliness'), we found those under 45 years old rated these as more important than those over 45. 
The paper concludes with a discussion of these results within the context of designing inclusive, digital systems that support a wide range of users.

\keywords{Artificial Intelligence \and Chatbots \and Conversational agents \and Data privacy \and Trust \and Personal information \and Human factors}
\end{abstract}

\section{Introduction}
Chatbots are software programs that can simulate a conversation in natural language, and are a promising technology for customer services in various contexts (e.g., finance, health care and tourism). Providing personalised experiences enables the provision of more effective user services. Personalisation is the ability to dynamically adapt functionality to an individual to better suit user needs \cite{fan2006personalization}. In a chatbot context, it is accomplished through the effective processing of a user's responses, and the identification and adaptation to information in disclosed during the conversation. Despite the advantages of chatbots, this process can often led to a tension between the requirements for service quality and the need for user privacy. These are both important topics within the context of human-computer interaction \cite{williams2019smartwatch}. 

Information disclosure to chatbots and factors that have an impact on it have been widely studied in the literature and the privacy concerns are often identified as a barrier to individual's disclosing information \cite{ghosh2020understanding,lee2013people,suganuma2020understanding,treiblmaier2013trust}. However, the main issues that lead to privacy concerns in chatbot interactions and the design practices that challenge data privacy principles, are often overlooked. Hence, in this study, we provide the perspective of British citizens privacy concerns surrounding the design practices of AI-based chatbots. 

We conducted an empirical study with 491 participants where four main challenges or ambiguities in agent design have been evaluated. These are, third-party access to personal information, inappropriate use of information once shared with chatbots, loss of control over personal data, and finally, ambiguities regarding the deletion of personal data. In addition, we investigated the factors that help to build trust in user/human-chatbot interaction. We investigated five factors for this purpose; the gender of the chatbot, use of a chatbot avatar, the quality of responses received from the chatbot, the friendliness of the chatbot, and the grammatical correctness of the language used by the agent. Our findings contribute to the literature by providing insights from a UK perspective on the main privacy concerns particular to chatbot design. These results can help to design user-centered solutions prioritising and respecting the privacy concerns of users. 

\section{Literature Review}

Folstad et al. \cite{folstad2018makes} categorised the factors perceived to affect trust in chatbots for customer service into two high-level groups; factors concerning the chatbots and factors concerning the service environment. Quality in interpretation of the user requests and advice in response to request were given as factors in the first category. They were followed by human-likeliness, self presentation (which describes the chatbot's communication of what it can do and its limitations) and professional appearance. Professional appearance in this context was defined as being thoughtfully developed and the chatbot providing grammatically correct responses. The authors reported three factors concerning the service environment; brand of the service provider, security and privacy aspects of the service, and the perceived risk associated with using the chatbot. The ability for the chatbot to correctly interpret the user requests and the  advice in the response were the most frequently reported factors identified by the participants.  Ischen et al. \cite{ischen2019privacy} investigated to what extent privacy concerns in chatbot interactions have an impact on users’ attitudes and their adherence to the recommendations provided by the chatbot. In addition, findings revealed that information disclosure is indeed influenced by privacy concerns. 

Even though the privacy concerns or perceived risk are a well-known factors that have impact on trust to chatbots, to the best of our knowledge, there is no study that investigates the issues behind perceived risk in user-chatbot interactions. However, some design practices in AI-based chatbots introduce several challenges to design privacy aware solutions. In their study, Saglam and Nurse identified open issues in agent design from a data privacy perspective where GDPR (General Data Protection Regulation) has been used as privacy regulation~\cite{saglam2020your}. The lack of algorithmic transparency, the difficulties in managing consent, and the difficulties in exercising the right to be forgotten, are some of the open issues raised. 

In this current study, bearing in mind the nature of chatbot technology and the challenges they provide for assuring data privacy, we prepared a survey where participants were asked to evaluate their concerns regarding deleting their personal information, third-party sharing, and losing control over their personal data. Our results contribute to formulate privacy concerns in chatbot design taking into account the design practices of this technology.

\section{Method} \label{section:Methodology}
We implemented a survey on the SurveyMonkey platform and asked participants to provide their opinions on two issues; potential risks that result in concerns in chatbot interaction and the factors that lead them to trust a chatbot. Before those questions, we posed questions to collect informed consent from the participants.  Demographic characteristics of the participants (age group, gender, and educational level) were also collected. The chatbot-centered questions used a 7-point Likert scale and asked the following questions:

\begin{itemize}
    \item To what extent do you agree or disagree with the following statements?
\begin{itemize}
  \item After using a chatbot, I would feel that others know about me more than I am comfortable with.
  \item After using a chatbot, I would worry that any personal information that I shared with it could be inappropriately used
  \item After using a chatbot, I believe that I would have control over who can get access to any personal information that I shared with it
  \item After using a chatbot, I would worry about how to delete any personal information that I shared with it
\end{itemize} 
\item How important are each of the following factors in determining whether you trust a chatbot or not?
\begin{itemize}
  \item The gender of the chatbot
  \item Whether the chatbot has an avatar (picture or visual depiction)
  \item The quality of responses received from the chatbot
  \item The friendliness of the chatbot
  \item The grammatical correctness of the language used.
\end{itemize}
\end{itemize}

\subsection{Participants}
This study's ultimate goal is to explore the perspective of British citizens on the privacy concerns of chatbot interactions. Therefore, we recruited participants using Prolific\footnote{https://www.prolific.co/}, which allowed us to reach a representative sample of UK citizens based on gender, age and ethnicity. We included six attention checking questions and excluded the participants who failed in more than one attention question. 

Within the 491 valid participants 49.7\% self-identified as male, and 10.4\% being aged between 18 and 24, 19.2\% between 25 and 34, 15.9\% between 25 and 44, 18.9\% between 45 and 54 and 35.6\% being 55 and over. When considering the highest-level of the participant's education 15.5\% achieved a GCSE-level of education\footnote{Typically taken at 15 years of age}, 28.1\% achieved an A-level or equivalent\footnote{A subject-based qualification between typically forming the period from leaving compulsory education to pre-university education}, 34.4\% achieved an undergraduate degree, 18.7\% a postgraduate degree and 3.3\% a doctorate.

\subsection{Analysis}
The analysis conducted for this study was a mixture of descriptive and quantitative statistics. Proportional-odds logistic regression models \cite{polr} were built to model the effects of age, gender and education. Ordinal regression, such as the proportional-odds logistic regression used in this study, is a common approach to modelling problems where the dependent variable (in this case, the Likert response) is ordinal. The model coefficients provide an insight into the effects of these variables on how participants rate their concerns or levels of trust.  

\section{Results} 
\label{section:Results}
\subsection{When to worry?}
\begin{figure}
    \centering
    \includegraphics[width=0.7\linewidth]{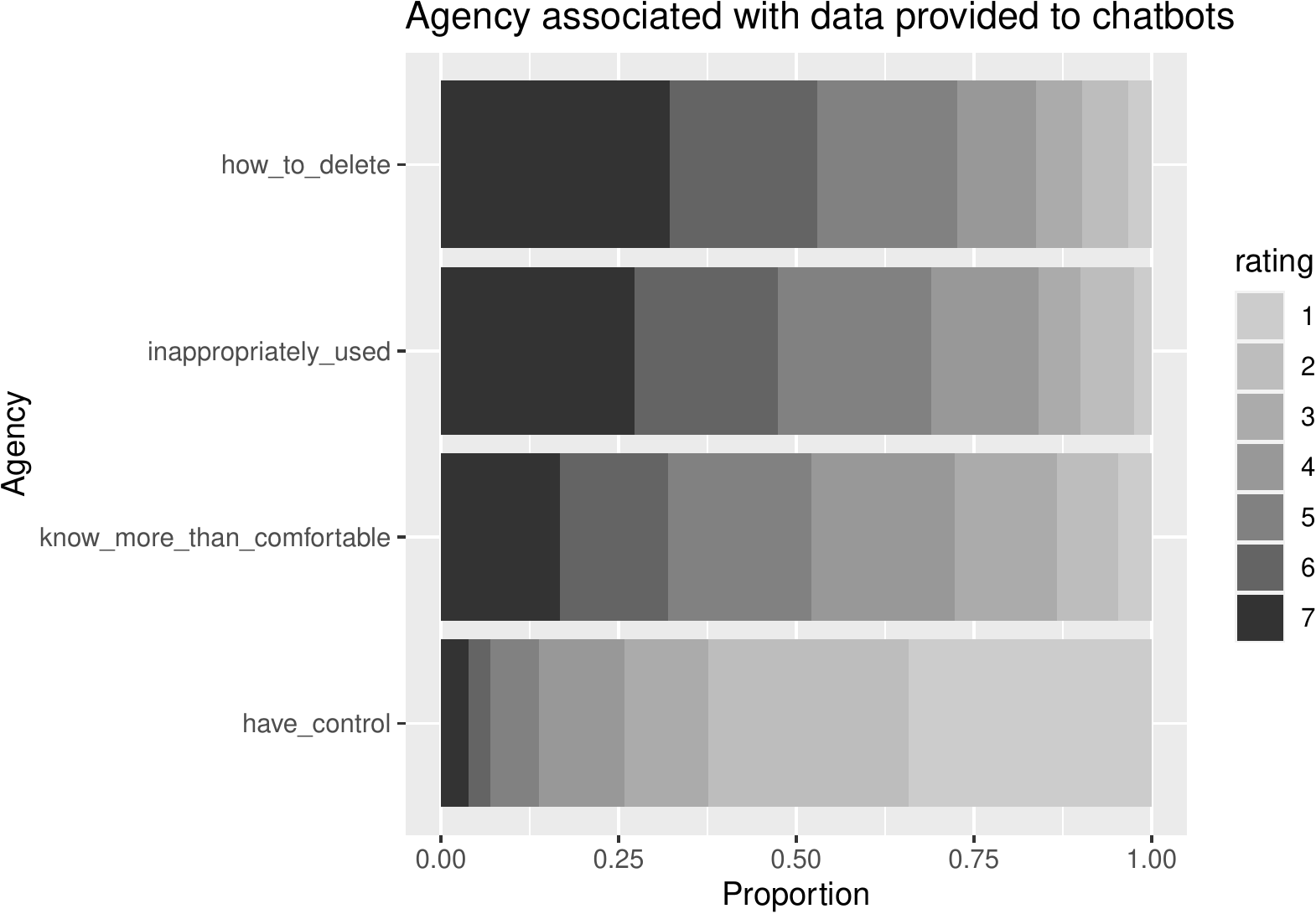}
    \caption{The responses from the data privacy concerns.}
    \label{fig:data_privacy_concern}
\end{figure}

We asked our participants to evaluate three main concerns: privacy concerns around others knowing more than they are comfortable with after chatbot interaction; worries around inappropriate use, and how to delete personal information. We also asked their beliefs on whether they would have control over who can access their personal information. The results from this are shown in Figure~\ref{fig:data_privacy_concern}.
 
Our findings reveal that UK citizens are most frequently concerned about how to delete personal information. Followed by concerns about whether the data would be inappropriately used.

\begin{figure}
    \centering
    \includegraphics[width=0.45\linewidth]{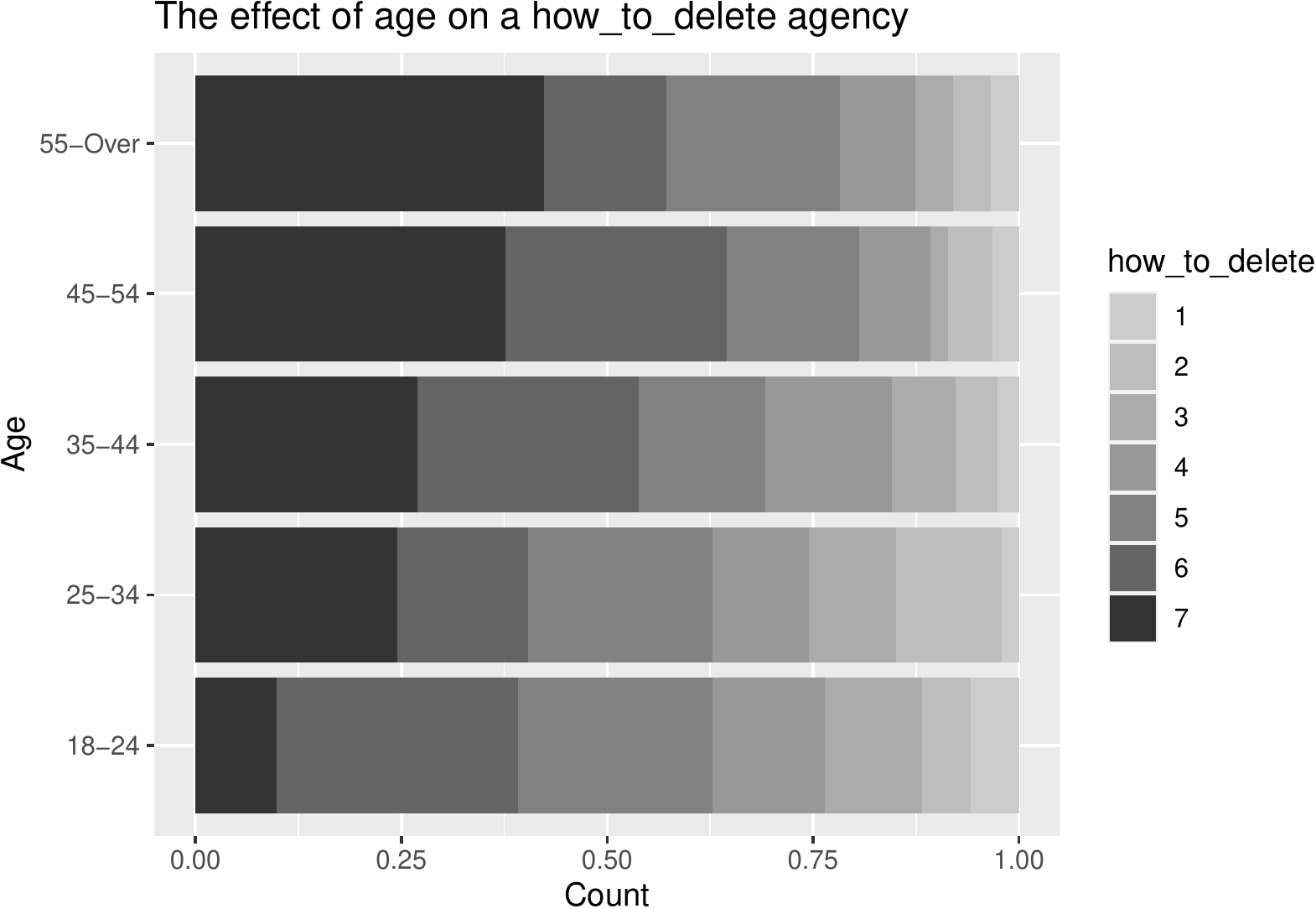}
    \includegraphics[width=0.45\linewidth]{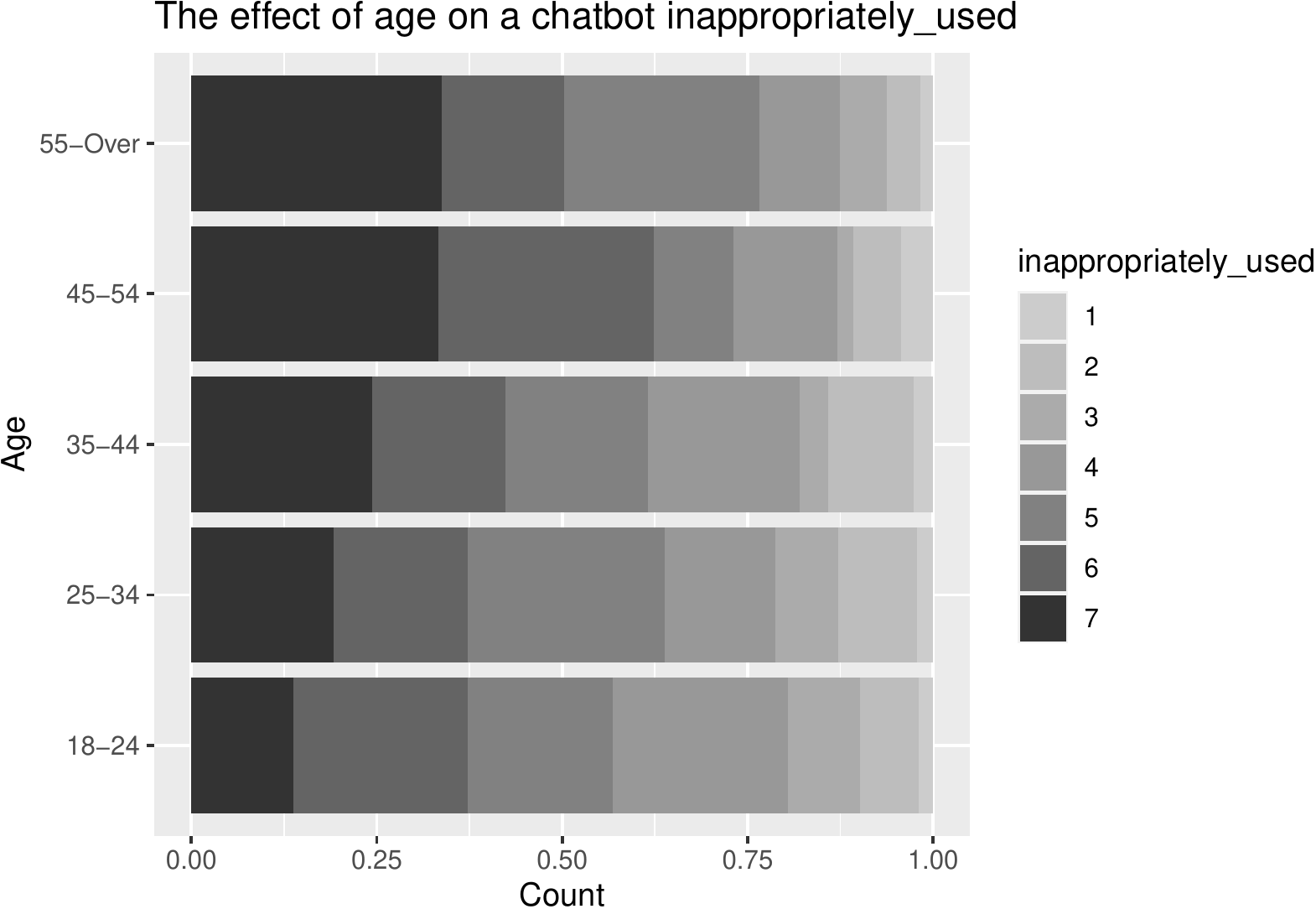} \\
    \vspace{0.6cm}
    \includegraphics[width=0.45\linewidth]{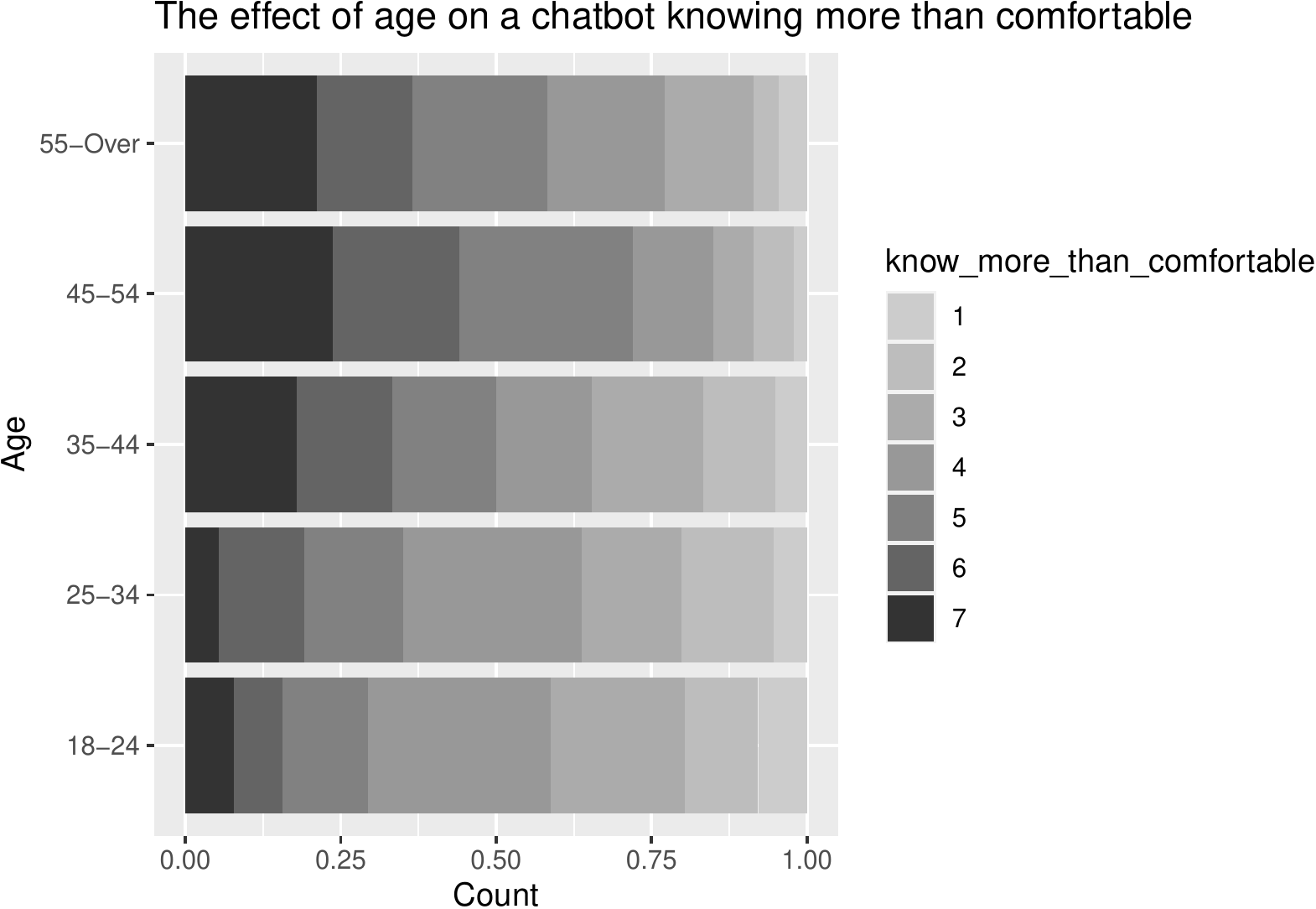}
    \caption{The effect of age on the data privacy concerns.}
    \label{fig:age_effects_concerns}
\end{figure}

When we perform a logistic regression analysis to examine the influence of user factors (age group, gender, education) on these concerns, we found no interaction from the gender or education variables. However, the proportional odds logistic regression model results revealed a significant age effect in the top three concerns. This effect is shown in Figure~\ref{fig:age_effects_concerns}. The logistic-regression models show statistical significance (at a 5\% significance level) for both the 45 to 55 age group and the over 55 age group.

\subsection{When to trust?}
\begin{figure}
    \centering
    \includegraphics[width=0.7\linewidth]{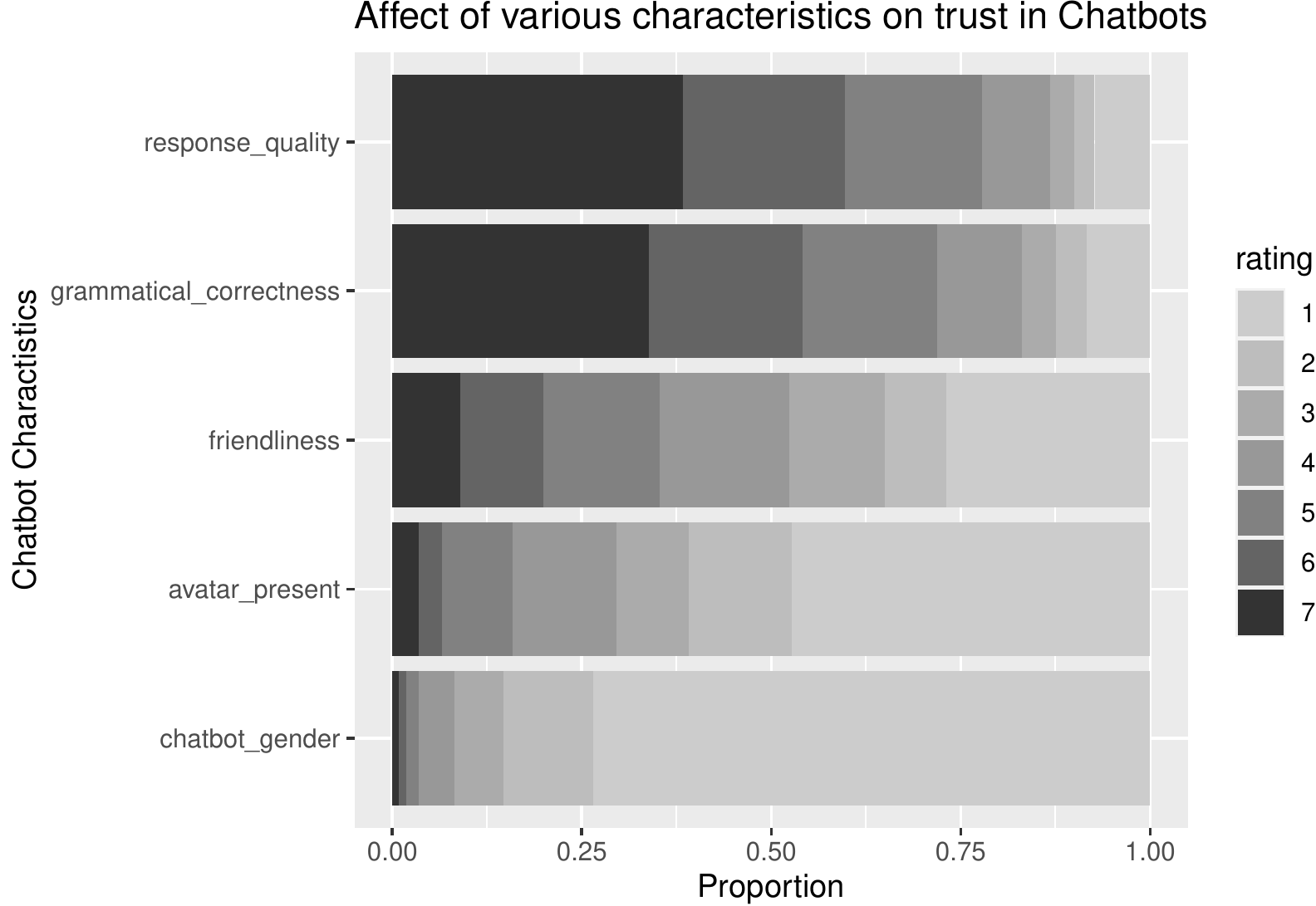}
    \caption{The responses when considering the characteristics which engender trust.}
    \label{fig:trust}
\end{figure}

When we analyse the responses given to our second question surrounding what engenders trust, we observe that there is a preference for `technical' quality in chatbots above other characteristics (see Figure~\ref{fig:trust}). This preference demonstrates the impact of the grammatical correctness and quality of the response over more `socially-driven' elements such as an avatar's presence, an interpreted gender or concepts such as `friendliness'.

\begin{figure}
    \centering
    \includegraphics[width=0.45\linewidth]{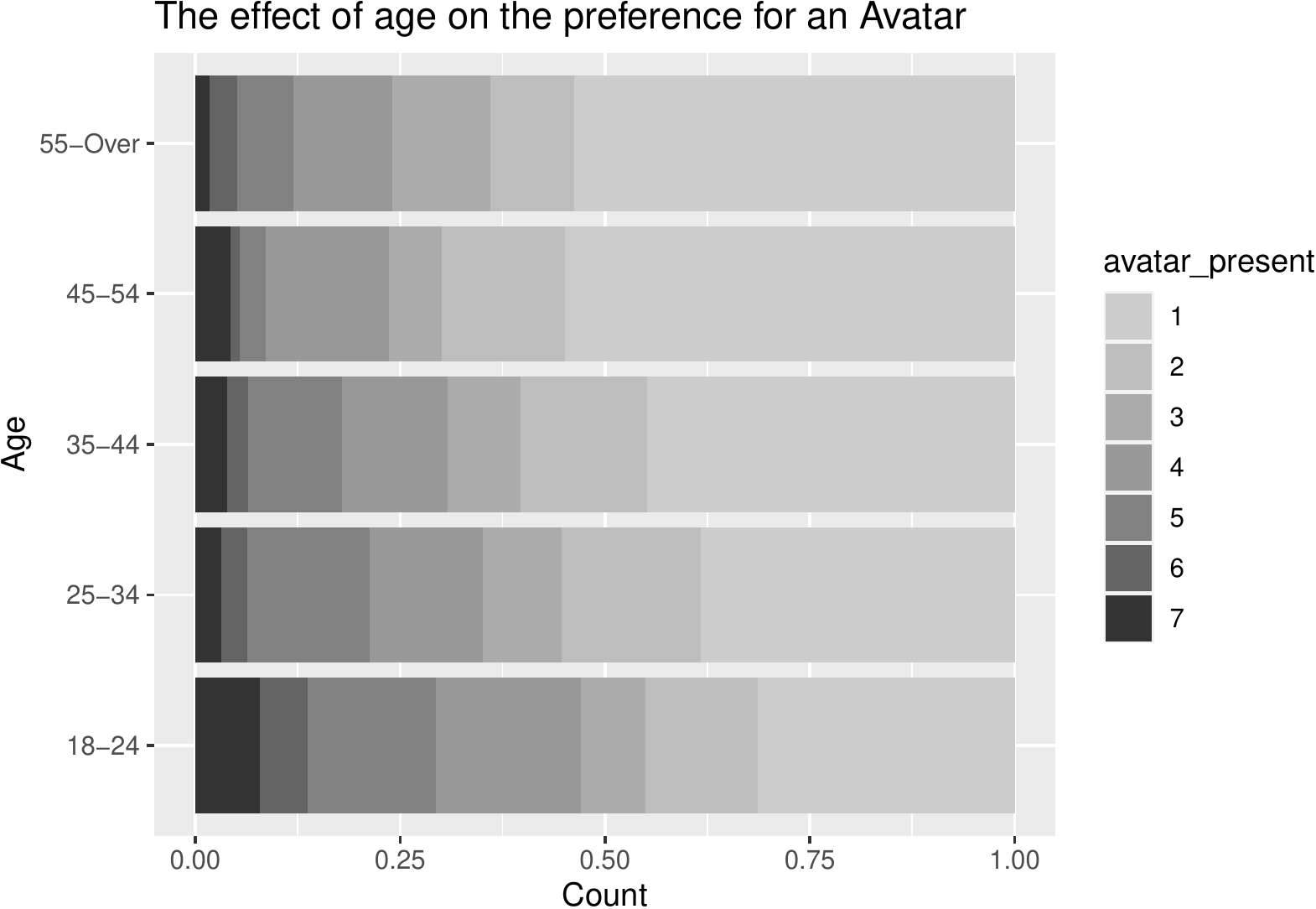}
    \includegraphics[width=0.45\linewidth]{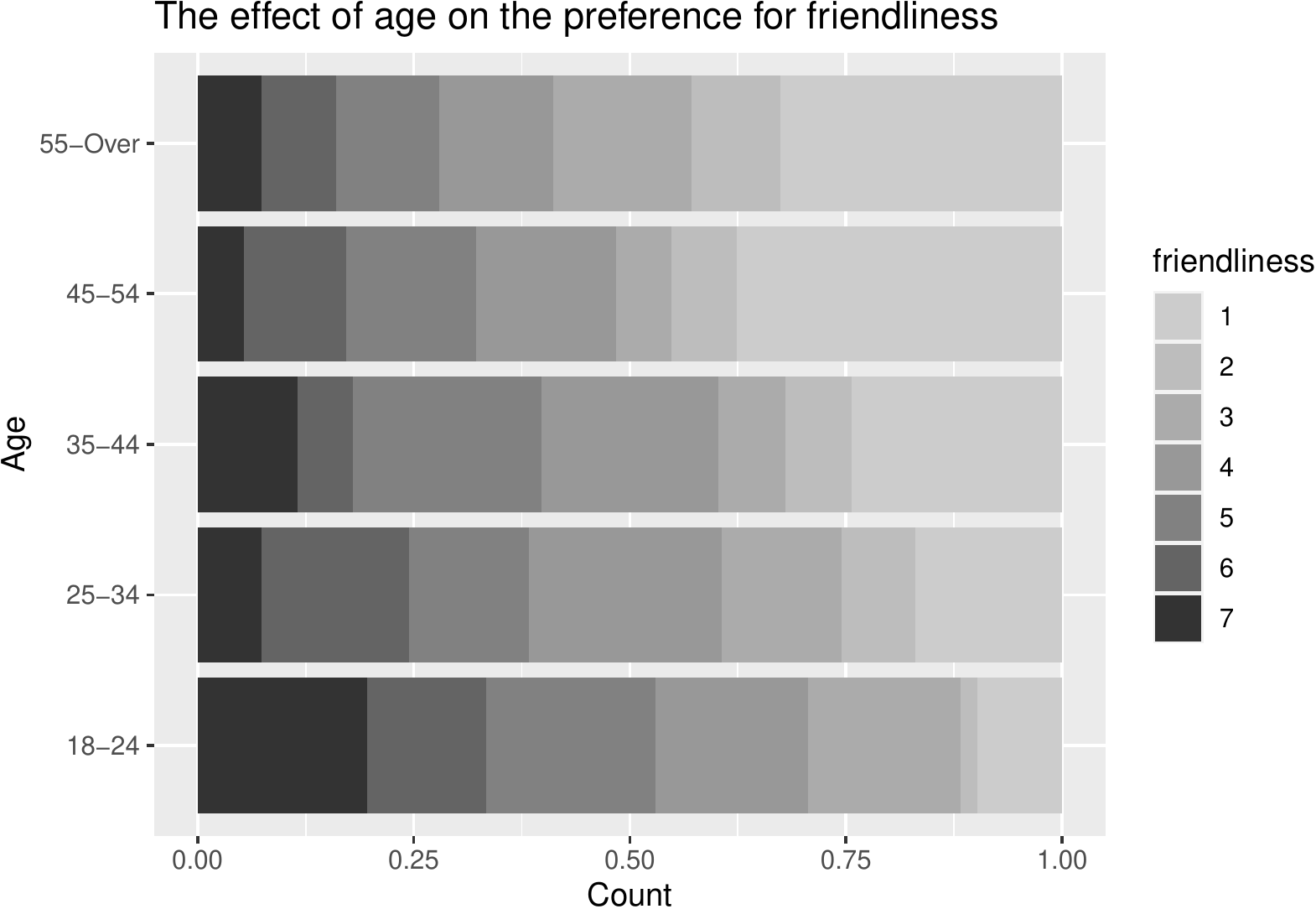}
    \caption{The effect of age on the chatbot characteristics that engender trust.}
    \label{fig:age_effects_characteristics}
\end{figure}

When we performed the same logistic regression analysis on this data, we again found no significant effect from either the participants' education level or the self-described gender. However, when we consider the participant's age, we again begin to see some interesting effects. When we consider the response quality, we can see no significant effect from the participants' age. There was also no age effect associated with the gender presented by the chatbot. The effect of age on the importance of the response's grammatical correctness was significant, although small for those in the highest age bracket, and the remaining age-groups exhibited no significant differences. The remaining two characteristics, whether an avatar was present and the `friendliness' of the avatar, did have an effect from the age of the participant. These characteristics are shown in Figure~\ref{fig:age_effects_characteristics} with those under 45  considering these characteristics more important when engendering trust with a chatbot.

\section{Discussion and Conclusion}
This study aimed to investigate users' privacy concerns and to understand features that are important to build trust in services mediated by chatbots.

The first emerging issue is the feeling of the loss of agency over data provided to a chatbot, with significant concerns over the ability to delete data and concerns around how the data would be used. The chatbot effectively decouples the individual disclosing the data from the final users of the data. This decoupling reduces the user's saliency of the value gained to the user from each disclosure; this has been shown to reduce the desire to disclose data \cite{lee2013people}. Hence, we should reduce this gap where possible and ensure users perceive they maintain agency over their data. The significant effect from the age of the respondent is an important observation. Suppose we are looking to build inclusive, digital systems that responsibly support citizens and user-bases. In that case, designers should be aware that these age groups may have more significant concerns in their use of chatbots. A system that explicitly addresses these concerns early in the interaction is likely to enable better disclosures and outcomes for all users.

The second emerging issue was the chatbot characteristics that engender trust in a chatbot. The factors which had the most significant effect were the response quality and the grammatical correctness of the responses. These factors represent the most salient cue to a user about the `competence' of the designer, builder and operator of the chatbot. It is perhaps not surprising that this engenders trust in the chatbot --- and it is noteworthy that this was consistent across all users independent of age, gender and education.

However, there was an age effect associated with the effect of two of the chatbot's characteristics. We identified that those under 45 were significantly more likely to use an avatar and the chatbot's `friendliness' when deciding whether to trust a chatbot. This separation between those over 45 and those under 45 appears several times in this dataset. We hypothesise that this relates to the Xennials, who had an analogue childhood but a digital adulthood. These individuals will have had the introduction to technology with a very un-human interface, which is likely to lower their expectations of machine interfaces. Those who have had a digital childhood and indeed come to computing where there are many more rich interfaces such as touchscreens and pervasive technology in the home perhaps set a higher expectation on these rich interactions. All of these issues are areas we are looking to explore in more depth.

A final comment would be orientated around the effect of participants age, particularly on the value placed on chatbots' characteristics. There are significant differences in the effect of avatars and the perceived `friendliness' of chatbots, particularly between those aged 18-24 and those aged over 45. If we are to base our understanding of how to design systems from an evidence-base driven by this younger cohort, we risk creating systems that are not effective for all ages. 

\section*{Acknowledgements}
This work is funded by the UK EPSRC `A Platform for Responsive Conversational Agents to Enhance Engagement and Disclosure (PRoCEED)' project (EP/S027211/1 and EP/S027297/1).

\bibliographystyle{splncs04}
\bibliography{main}
\end{document}